\documentstyle[preprint,aps]{revtex}
\tightenlines

\newfont{\lfig}{cmr10 scaled\magstep0} 
\begin{document}
\draft
\title{Chiral Dynamics and the Low Energy Kaon-Nucleon Interaction$^*$}
\author{N. Kaiser, P.B. Siegel$^{**}$, and W. Weise}
\address{Physik Department, Technische Universit\"{a}t M\"{u}nchen\\
   Institut f\"{u}r Theoretische Physik, D-85747 Garching, Germany}

\bigskip

\bigskip

\maketitle
\begin{abstract}
We examine the meson-baryon interaction in the strangeness $S=-1$ sector using
an effective chiral Lagrangian.  Potentials are derived from this Lagrangian
and used in a coupled-channel calculation of the low energy observables.  The
potentials are constructed such that in the Born approximation the s-wave
scattering amplitude is the same as that given by the effective chiral
Lagrangian,
up to order $q^2$.  Comparison is made with the available low energy hadronic
data of the coupled $K^-p, \Sigma \pi, \Lambda \pi$ system, which includes the
$\Lambda (1405)$ resonance, $K^-p$ elastic and inelastic scattering,  and the
threshold branching ratios of the $K^-p$ decay.  Good fits to the experimental
data and estimates of previously unknown Lagrangian parameters are obtained.

\end{abstract}

\vspace{3in}

\noindent $^*${\it Work supported in part by BMBF and GSI}

\noindent $^{**}${\it On sabbatical leave from California State Polytechnic
University, Pomona CA 91768}

\newpage

\section{Introduction}

The effective chiral Lagrangian including baryons, which corresponds to an
expansion in increasing powers of derivatives of the meson fields and quark
masses, has been successful in understanding many properties of the
meson-baryon system at low energies \cite{bernard1}.
Since the pion mass is small, properties of the pion-nucleon system are well
described close to threshold.   Recently, the $SU(3)$ chiral Lagrangian has
also been applied to the $K^+N$ system \cite{lee,brown}.  In both the $\pi N$
and $K^+ N$ cases, the low energy (s-wave) interaction is relatively weak and
the leading term of the Lagrangian (linear in the meson four-momentum $q$) is
the main one.  In this article we examine the $\overline{K}N$ system within the
context of the effective chiral Lagrangian.  In contrast to the $\pi N$ and
$K^+ N$ cases, the $\overline{K}N$ system is a strongly interacting
multichannel system with an s-wave resonance, the $\Lambda (1405)$, just below
the $K^-p$ threshold.  Terms of higher order in $q$ will therefore play a more
pronounced role in the interaction.  In fact the dynamical generation of such a
resonance necessarily leads beyond any finite order expansion of chiral
perturbation theory.

The motivation for nevertheless applying the effective chiral Lagrangian to the
$K^-N$
system is two-fold: first, to explore whether it provides driving terms with
sufficiently strong attraction to generate a resonance at the right energy;
secondly, to test the theory in the $S=-1$ sector and to obtain
information about the parameters of the Lagrangian not accessible in the $\pi
N$ or $K^+ N$ interactions.  At low energies, the $K^-$ and proton can scatter
into six different $\overline{K} N$ and pion-hyperon channels, and the data
place constraints on the relative interaction strengths among the different
channels.  Since each of the different
terms of the chiral Lagrangian has its own particular SU(3) structure, this
six-channel system is a good testing ground for the theory and enables one to
obtain estimates of the Lagrangian parameters.

The presence of the $\Lambda (1405)$ resonance requires that one needs to go to
infinite order in a certain class of rescattering diagrams. Thus to build up
the resonance we are led to constructing a potential model to handle the
calculation.  A potential model was previously applied to the
$\overline{K} N$ system \cite{dalitz}, and the
approximations necessarily involved in using this method are
expected to be quite reasonable, since the resonance dominates the
interaction. In order to connect the chiral Lagrangian to the potential model
approach, we construct a meson-baryon potential which in the Born approximation
has the same s-wave scattering amplitude as the effective chiral Lagrangian, up
to order $q^2$.  Two different forms of the potential, one local and one
separable, are investigated.  This potential is iterated in a
Lippmann-Schwinger equation, and the free parameters of the Lagrangian are
adjusted to fit a wide variety of low-energy data of the kaon-baryon system.
An additional constraint is placed on the parameters so that the isospin even
$\pi N$ s-wave scattering length is consistent with its experimental value.  We
find that parameters from Lagrangian terms up to order $q^2$ are sufficient to
fit the data.  In particular, the observed properties of the $\Lambda (1405)$
resonance are reproduced remarkably well.  Furthermore as a consistency check,
we calculate the $KN$ s-wave phase shift (strangeness $S=+1$ sector) from the
parameters obtained in fitting the $\overline{K} N$ system.

In the second section we describe the effective chiral Lagrangian and the
potential model, and in the third section we discuss our results.  In each
section we consider interaction terms of order $q$ and then of order $q^2$,
where $q$ stands for the meson four-momentum.  In the final section we discuss
the limitations and some qualitative aspects of the calculation.

\section{Effective Chiral Lagrangian}

The effective low-energy Lagrangian density for interacting systems of
pseudoscalar mesons and baryons can be written generally as
\cite{gasser,krause}

\begin{equation}
  {\cal L} = {\cal L}^{(1)} + {\cal L}^{(2)} + \cdots
\end{equation}

\noindent corresponding to increasing number of derivatives (external momenta)
and quark masses.

\subsection{Leading Order Term}

In the relativistic formalism the leading order term \cite{gasser,krause} is
given by:

\begin{equation}
  {\cal L}^{(1)} = Tr( \overline{\Psi}_B(i \gamma_\mu D^\mu - M_0) \Psi_B ) +
           F \, Tr( \overline{\Psi}_B \gamma_\mu \gamma_5 [ A^\mu ,\Psi_B]) +
           D \, Tr( \overline{\Psi}_B \gamma_\mu \gamma_5 \{ A^\mu ,\Psi_B \})
\end{equation}

\noindent with the chiral covariant derivative

\begin{equation}
   D^\mu \Psi_B = \partial^\mu \Psi_B + [\Gamma^\mu,\Psi_B],
\end{equation}

and

\begin{equation}
   \Gamma^\mu = {1 \over {8f^2}} [\phi,\partial^\mu \phi] + O(\phi^4).
\end{equation}

\noindent The matrices $\Psi_B$ and $\phi$ represent the octet baryon Dirac
fields and the octet
pseudoscalar meson fields, respectively (see Appendix I), $f \simeq 90$
MeV is the pseudoscalar meson decay constant and $M_0$ the baryon mass in the
chiral limit \cite{leutwyler}.  The
constants $F$ and $D$ are the $SU(3)$ axial vector couplings, whose accepted
values are $F \simeq 0.5$ and $D \simeq 0.75$ \cite{bourquin}, together with
$g_A = F+D$.
Through Goldberger-Treiman relations they determine the pseudovector meson
baryon coupling strengths.

The interaction part involving the covariant derivative can be expanded, and to
leading order in the external meson four-momentum $q$ is given by

\begin{equation}
  {\cal L}^{(1)}_{int} = {i \over {8f^2}} Tr(\overline{B}[[\phi,
                                       \partial_0 \phi],B])
\end{equation}

\noindent where the field $B$ is the "large component" of the Dirac spinor
$\Psi_B$, defined via  $\Psi_B = e^{-iM_0 v \cdot x} (B + O(1/M_0))$
\cite{jenkins,bernard}.  The
four-velocity $v$ permits fixing a particular
reference frame which we choose to be the meson-baryon center of mass system.
An observer in that frame is characterized by $v^\mu = (1,0,0,0)$.    This part
of the interaction, Eq.(5) is the current algebra, or
Weinberg-Tomozawa, term \cite{weinberg}.  The tree approximation of this term
agrees with the
experimental values of the s-wave scattering length to within $15\%$ for $\pi
N$ and to within $50 \%$ for the $K^+ N$ system.  In the $K^+ p$ channel this
Weinberg-Tomozawa term is repulsive.  The Lagrangian also applies to the $K^-
N$ interaction.  Here, the current algebra term produces an attractive
interaction.  The $\Lambda (1405)$ resonance just below the $K^- p$ threshold
could be resulting from this attractive term.  We investigate this possibility
in the next section.

\subsection{Order $q^2$ terms}

In addition to the order $q$ terms mentioned above, the tree graphs of the
relativistic Lagrangian ${\cal L}^{(1)}$ give rise to s-wave meson-baryon
amplitudes at order $q^2$ (and higher).  These order $q^2$ terms are
relativistic ($1/M_0$)
corrections from the covariant derivative piece and the Born graphs involving
the axial coupling terms ($F$ and $D$).  In addition to these order $q^2$
pieces,
we have to consider the most general form of the chiral effective Lagrangian
${\cal L}^{(2)}$.  The complete relativistic version involving all terms
allowed by
chiral symmetry can be
found in \cite{krause}.  In the heavy baryon formalism its form is simpler, and
the relevant s-wave terms are \cite{brown}:

\begin{eqnarray}
{\cal L}^{(2)} &=& b_D Tr(\overline{B} \lbrace \chi_+ , B \rbrace)+b_F
 Tr(\overline{B} \lbrack \chi_+ , B \rbrack)  + b_0 Tr(\overline{B} B)
 Tr(\chi_+) \nonumber \\
        & & + d_D Tr(\overline{B} \lbrace (A^2 + (v \cdot A)^2),B \rbrace) +
        d_F Tr(\overline{B} \lbrack (A^2 + (v \cdot A)^2),B \rbrack) \nonumber
\\
        & & +d_0 Tr(\overline{B} B) Tr(A^2 + (v \cdot A)^2)   \\
    & & + d_1 (Tr(\overline{B} A_\mu) Tr(A^\mu B) + Tr(\overline{B} (v \cdot
A)) Tr((v \cdot A) B) ) \nonumber \\
    & &  + d_2 Tr(\overline{B} (A_\mu B A^\mu + (v \cdot A) B (v \cdot A)))
                  + \cdots \nonumber
\end{eqnarray}

\noindent where

\begin{equation}
  \chi_+ = - {1 \over {4 f^2 }} \lbrace \phi, \lbrace \phi, \chi\rbrace \rbrace
{}.
\end{equation}

The matrix $\chi$ is proportional to the quark mass matrix
and has only diagonal elements (See Appendix I).  Using the Gell-Mann, Oakes,
Renner relation in the isospin limit for the pseudoscalar mesons, the diagonal
elements of $\chi$ are $(m^2_\pi , m^2_\pi , 2m^2_K - m^2_\pi )$.  The axial
matrix operator $A$ is given by

\begin{equation}
   A_\mu = - {1 \over 2 f} \partial_\mu \phi + O(\phi^3) .
\end{equation}

All contributions of ${\cal L}^{(2)}$ at tree level are of order $q^2$.
The next order $q^3$ involves loops. The amplitudes for all channel couplings
up to order $q^2$, which also include the relativistic corrections, are given
in Appendix II.

The constants in front of the various terms are not determined
from chiral symmetry, but need to be determined from experiment.  The
first two, $b_D$ and $b_F$, can be determined from the baryon mass splittings
since to order $q^2$:

$$
    m_\Sigma - m_\Lambda = {16 \over 3} b_D(m^2_K - m^2_\pi)
$$

\noindent and

\begin{equation}
  m_\Xi - m_N = 8b_F(m^2_\pi - m^2_K).
\end{equation}

\noindent One obtains $b_D = +0.066$ GeV$^{-1}$ and $b_F = -0.213$
GeV$^{-1}$.   Data restricts the range for the constant $b_0$, since it is
related to the $\pi N$ sigma term.  To order $q^2$ we have:

\begin{equation}
    \sigma_{\pi N} = -2 m^2_\pi (b_D + b_F + 2 b_0),
\end{equation}

\noindent and the strangeness content of the proton to this order is given by:

\begin{equation}
    {{<p|\overline{s}s|p>} \over  {<p|\overline{u}u + \overline{d}d |p>}} =
                      {{b_0+b_D-b_F} \over {2b_0+b_D+b_F}}.
\end{equation}

\noindent Thus, $b_0$ cannot be greater than $-0.28$ GeV$^{-1}$ since
otherwise the strangeness content of the proton would be negative.  We also did
not allow $b_0$ to be less than $-0.52$ GeV$^{-1}$, which keeps $\sigma_{\pi
N}$ less than about $45$ MeV \cite{sigma}.  Data from interactions among the
meson-baryon octet (i.e. the $K^-N$ interaction) gives information regarding
the double-derivative parameters, the $d$-terms in Eq.(6).  The notation in
Eq. 6 was adopted since if $\chi$ is proportional to the unit matrix
(i.e. $m_u=m_d=m_s$), terms with the same subscripts have the same relative
SU(3) strengths.

\subsection{The Pseudo-Potential}

The motivation for studying the $K^-N$ system is to see how well the effective
chiral Lagrangian describes the experimental data in this strangeness sector,
and to obtain estimates for the "$d$" parameters of Eq.(6).  Since the
interaction supposedly produces a resonance just below the $K^-p$ threshold, a
certain class of diagrams needs to be summed to infinite order to produce a
pole
at the appropriate energy.  Thus, the chiral perturbation expansion for the
scattering amplitude is not feasible, and we adopt a potential model approach.
We connect the two formalisms in the following way:  {\it the pseudo-potentials
are constructed such that up to order $q^2$ they have, in the Born
approximation, the same s-wave
scattering amplitude as the chiral Lagrangian}.  In the last
section we discuss some of the limitations of using this potential approach.

The above prescription for obtaining a potential from the chiral Lagrangian
gives a unique zero-range potential of the form:

\begin{equation}
  V_{ij}(\vec{r}) = {{C_{ij} } \over {2 f^2 }} \sqrt{{{M_i M_j}
       \over {s \omega_i \omega_j }}}  \delta^3 (\vec{r})
\end{equation}

\noindent where the $i$ and $j$ label the six meson-baryon channels \\
($\pi^+ \Sigma^-,\; \pi^0 \Sigma^0, \; \pi^- \Sigma^+, \; \pi^0 \Lambda, \; K^-
p, \; \overline{K^0} n$).  Here $\sqrt{s}$ is the total center-of-mass energy
of the system, $M_i$ is the mass of the baryon in channel $i$, and $\omega_i$
is
the reduced energy of channel $i$ in the center-of-mass frame.  Explicitly,

\begin{equation}
   \omega_i = {E_i \sqrt{E_i^2+M_i^2-m_i^2} \over E_i+
                          \sqrt{E_i^2+M_i^2-m_i^2}} ,
\end{equation}

\noindent where $m_i$ is the
mass and $E_i$ is the energy of the meson in channel $i$.  The $C_{ij}$ are
determined directly from the effective chiral Lagrangian and
are given in Appendix II.   The kinematic factor $\sqrt{M_i M_j / (s \omega_i
\omega_j)}$ is needed to obtain the proper relativistic flux factor in the
differential cross section.  The differential cross section from initial
channel $j$ to final channel $i$ in the Born approximation is:

\begin{eqnarray}
 \biggl( {{d \sigma_{ij}} \over {d \Omega_{cm}}} \biggr)_{Born}
                     &=& {{k_i} \over {k_j}}
                        \bigg|{{\sqrt{\omega_i \omega_j}} \over {2 \pi}}
            \int e^{-i(\vec{k_j}-\vec{k_i})\cdot
               \vec{r}} V_{ij}(\vec{r}) d^3 r \bigg|^2
                     \nonumber \\
                &=& {k_i \over k_j} {{M_i M_j} \over {16 \pi^2 s}}
                        {{|C_{ij}|^2} \over {f^4}}
\end{eqnarray}

\noindent where we have used Eq.(12) in the last step, and  $k_i$ stands for
the
meson momentum in the center of mass frame.

In order to dynamically produce the $\Lambda (1405)$ resonance, the potential
must be iterated to infinite order.  Here we use a
Lippmann-Schwinger equation.  To obtain a finite result, a cut-off or
inverse range scale, denoted by $\alpha$, needs to be introduced in the
potential.  We examine two ways of
parameterizing the finite range of the potential while keeping the Born term
the
same: a local potential and one separable in the incoming and outgoing relative
momenta.  For the local potential between channel i and j we use a Yukawa form:

\begin{equation}
   V_{ij}(\vec{r}) = {{C_{ij} \alpha^2_{ij}} \over {8 \pi f^2}} \sqrt{{{M_i
M_j}  \over  {s \omega_i \omega_j }}}  {{e^{-\alpha_{ij} r}} \over {r}}
\end{equation}

\noindent where $r = |\vec{r}|$.  In our analysis, we found a
satisfactory solution for the local potential using only a single common range
$\alpha=412$ MeV for all channels.  For the Yukawa potential the parameter
$\alpha$ can be interpreted as an average "effective mass" that represents the
spectrum of exchanged particles in the t-channel mediating the interaction.
The value of $\alpha$ found here for the $I=0$ channel is reminiscent of
typical two-pion ranges.

For comparison we also examine a separable potential of the momentum space
form:

\begin{equation}
   V_{ij}(k_i,k_j) = {{C_{ij}} \over {4 \pi^2 f^2}}  \sqrt{{{M_i M_j} \over
              {s \omega_i \omega_j }}}  {{\alpha^2_i} \over
       {\alpha^2_i+k^2_i}} {{\alpha^2_j} \over {\alpha^2_j+k^2_j}}
\end{equation}

\noindent where $k_i$ is the center-of-mass momentum for channel i.
The potentials are inserted into a Lippmann-Schwinger equation.  For the local
potential, this equation is solved in coordinate space:

\begin{equation}
  \nabla^2 \psi_i + k^2_i \psi_i = 2 \omega_i \sum_j V_{ij} \psi_j.
\end{equation}

\noindent For the separable potential we use the Lippmann-Schwinger equation
(for the s-wave) in momentum space:

\begin{equation}
    T_{ij}(k_i,k_j) = V_{ij}(k_i,k_j) +
           \sum_n  \int^\infty_0 {{q^2 dq \, 2 \omega_n V_{in}(k_i,q)
           T_{nj}(q,k_j)} \over {q^2 - k^2_n + i \epsilon }},
\end{equation}

\noindent which can then be solved analytically.  The T-matrix resulting from
this equation is related to the nonrelativistic scattering amplitude  $f_{ij} =
- \pi \sqrt{\omega_i \omega_j} T_{ij}$.  The total s-wave
cross section for the channel $j \rightarrow i$ is thus given by

\begin{equation}
   \sigma_{ij} = 4 \pi^3 {{k_j} \over {k_i}} \omega_i \omega_j |T_{ij}|^2 .
\end{equation}

\noindent Keeping only the Born term and using a zero range potential,
i.e. $\alpha_i = \infty$, this cross section becomes:

\begin{equation}
   \sigma^{Born}_{ij} = 4 \pi {k_j \over k_i}
                 {{M_i M_j} \over {16 \pi^2 s}} {{|C_{ij}|^2} \over
                     {f^4}}
\end{equation}

\noindent which agrees with the result of Eq.(14).

For our applications, the local potential is the more physical one, since it
permits an interpretation in terms of t-channel exchange processes.   For
example, vector meson exchange has the same basic SU(3) structure as the
current
algebra term.  However, it is useful also to examine the separable potential
for a qualitative analysis and to check for model dependencies.  As will be
shown in the next section, the results of our analysis and the values of the
"$d$" parameters turn out to be very similar for the two different potentials.
Hence the physics described here is not sensitive to the specific form of the
potential.

\section{Results}

Since the Weinberg-Tomozawa (current algebra) term is known to be the largest
one in the s-wave $\pi N$ interaction, it is instructive to examine how well
the interaction due to this term alone reproduces the available data in the
$S=-1$
sector.  So we start by examining the current algebra term by itself first, and
then add terms involving order $q^2$ in the chiral Lagrangian.

\subsection{Current Algebra term at lowest order}

The potential for this term is obtained by setting all the "$b$" and "$d$"
parameters to zero in $C_{ij}$ and neglecting the relativistic
corrections in Eq.(2).  The only free parameters are
the inverse ranges $\alpha_i$.  First we investigate if the current algebra
term alone will produce a resonance or quasi-bound state.  To do this, we
use a common range $\alpha$ for all channels.  The results are shown in Table
I, where we list the energy of the quasi-bound state formed for different
parameters $\alpha$.  As can be seen, if the range parameter of the interaction
for the local potential is larger than 300 MeV then a
quasi-bound state is indeed formed below the $K^-p$ threshold.  The range of
the interaction is roughly $0.7$ fm.  For the separable potential, the same
resonance below the $K^-p$ threshold is produced for a value of $\alpha$
greater than 400 MeV.

Thus, a potential derived from chiral dynamics with interaction ranges
commensurate with the meson-baryon system {\it necessarily produces a
quasi-bound state or resonance below or near the $K^-p$ threshold}.

How well does the current algebra piece predict the other low energy data?
First we consider the scattering data.  We have performed a $\chi^2$ analysis
on the low energy scattering data \cite{ciborowski,evans,humphrey,kim,sakitt}
as follows:  first a common range
$\alpha$ was picked, then the value of the decay constant $f$ was determined to
produce a resonance at 1405 MeV.  Then the total cross sections
were calculated, compared with the available experimental data, and a
chi-square per data point, $\chi^2/N$, was determined.  The parameter
$\alpha$ was varied until $\chi^2/N$ was minimized.  Note that since the
energy is low, only the $l=0$ partial wave needs to be considered.   The
results are shown in Table II.  We see that for both potentials, a fairly
good fit, $\chi^2/N =2.0$, to the scattering data is possible.  For the
local potential, the mimimum came out to be at $f = 90$ MeV, which almost
coincides with the pseudoscalar meson decay constant in the chiral limit.
It is quite remarkable that the current algebra term alone, when iterated using
a common off-shell range parameter, gives agreement with six scattering
channels
and produces an I=0 resonance at 1405 MeV.  For the separable potential, a
$\chi^2$ minimum was found at $f = 110$ MeV.

For the $K^+N \; S=+1$ sector, the un-iterated current algebra term gives zero
for the $I=0$ and $-0.585$ fm for the $I=1$ s-wave scattering length.
Experimentally, the $I=0$ scattering length is indeed very small.  To obtain
the experimental value of $-0.32$ fm for the $I=1$ scattering length, range
parameters $\alpha = 600$ MeV for the local and $\alpha = 557$ MeV for the
separable potential were needed.

In examining the threshold branching ratios, the situation is not so good for
the current algebra term alone.  The hadronic threshold branching ratios are
\cite{nowak,tovee}:

\begin{eqnarray}
    \gamma &=& {{\Gamma(K^-p \rightarrow \pi^+ \Sigma^-)} \over
       {\Gamma(K^-p \rightarrow \pi^- \Sigma^+)}} = 2.36 \pm 0.04  \nonumber \\
    R_c &=& {{\Gamma(K^-p \rightarrow charged ~particles)} \over
       {\Gamma(K^-p \rightarrow all)}} = 0.664 \pm 0.011  \nonumber \\
    R_n &=& {{\Gamma(K^-p \rightarrow \pi^0 \Lambda)} \over
       {\Gamma(K^-p \rightarrow ~all~ neutral~ states)}} = 0.189 \pm 0.015
\end{eqnarray}

\noindent They place tight constraints on the relative potential strengths of
the $\overline{K}N$ system.   Two of the  branching ratios are in fair
agreement with experiment, however, the strangeness plus double charge exchange
branching ratio $\gamma$ is far off.  Experimentally, the channel $\pi^+
\Sigma^-$ is produced 2.4 times as often as $\pi^- \Sigma^+$ in the decay at
threshold.  Note that the current algebra piece in lowest order gives zero for
the $K^-p \rightarrow \pi^+ \Sigma^-$ interaction.  In terms of chiral power
counting
the numerator goes as $q^4$ and the denominator as $q^2$, and the ratio
$\gamma$ is formally suppressed.  Experimentally this is not the case, which
demonstrates the limits of using a chiral perturbation expansion for the low
energy $\overline{K}N$ system.

We let the ranges vary for each channel to see if it was possible to fit both
the scattering data and the three threshold branching ratios.  No fit was
found.  Next we examine if it is possible to match the data by including the
terms of order $q^2$.

\subsection{Order $q^2$ Terms}

One cannot vary $b_0$ and the "$d$" parameters arbitrarily to fit the
$\overline{K}N$ data, since the chiral Lagrangian applies to all meson-baryon
interactions.  For two s-wave  scattering lengths ($I=0 \; KN$ and isospin-even
$\pi N$) the current algebra term gives zero and the leading term is order
$q^2$.  Experimentally, these scattering lengths are small: $a^+_{\pi N} =
-(0.012 \pm 0.006)$ fm \cite{koch} and $a^0_{KN} = -(0.02 \pm 0.04)$ fm
\cite{dover}.  Since $b_0$ and
the "$d$" parameters are the leading terms for these two scattering lengths, we
restrict their values in the search to be compatible with the experimental
results.  The $\pi N$ isospin-even and $I=0 \; KN$ scattering lengths are given
in Appendix II to order $q^3$ and $q^2$ respectively.  For the $I=1 \; KN$
scattering length, the range in the potential can be adjusted to fit the
experimental value of $-0.32$ fm.

The search was done as follows: First the nine parameters $b_0$, $d_0$, $d_D$,
$d_F$, $d_1$, $d_2$, and the range(s) were chosen.  Then a value for $f$
was determined which produced a resonance at 1405 MeV.  The total $\chi^2$ for
the scattering data to the six channels, the three threshold branching ratios,
and the scattering lengths $a^+_{\pi N}$ and $a^0_{KN}$ was calculated.  In
order to have the pion decay constant $f$ be near 93 MeV, the term $((f -
93)/0.02)^2$ was added to the $\chi^2$ function.  The nine
parameters were varied until the $\chi^2$ was minimized.  The search was
carried out using MINUIT \cite{minuit}, and some parameters were restricted in
the search.  The range parameters $\alpha$ were limited to vary between 300 and
900 MeV. The parameter $b_0$ was allowed to vary between $-0.28$ and $-0.52$
GeV$^{-1}$, which corresponds to $\sigma_{\pi N}$ less than $45$ MeV and
$<p|\overline{s} s|p>$ greater than zero.  It is interesting that for the local
potential all the best-fit parameters were within the ranges set.  A value of
$b_0=-0.493$ GeV$^{-1}$ corresponds to $\sigma_{\pi N} = 44$ MeV, which turns
out to coincide with the empirical analysis \cite{sigma}.

\section{Discussion}

Satisfactory fits were found for both the local and separable potentials using
appropriate range parametrizations.  For the separable potential, the best
fit was found using three different ranges for the $\pi \Sigma$, $\pi
\Lambda$, and $\overline{K} N$
channels.  For the local potential a good fit is obtained using only one
common range for all channels.  The values of the "best fit" parameters for
the two cases are listed in Table II.  In Table III we list
the $\chi^2 /N$ for the scattering data along with the values for the
threshold branching ratios. In Figs. 2-6 we plot the total cross sections to
the various channels, as well as
the $\Sigma \pi$ mass spectrum for the two fits.  The total $\chi^2$ can be
slightly improved for the local potential by using three different ranges as in
the separable case.  However,  the Lagrangian parameters do not change
significantly.  Thus, we focus our attention on the
single range fit, since it reproduces the data well and has the least
number of parameters.  Also, the preferred value $\alpha = 412$ MeV lies
between the mass of a vector meson and that of two pions.  This
value is in line with the physics of the process, since such
t-channel exchanges are believed to dominate the interaction.

As can be seen from Table II, the
Lagrangian parameters turn out to be very similar for both the local and
separable potential fits.  This is probably because the range of energies
investigated here is small, and thus the results are not sensitive to the form
of the potentials.  If other potential forms were used, similar values for the
Lagrangian parameters would probably be obtained.   Thus, the most important
parameters in the fit are "$b_0$" and the "$d$'s".  To our knowledge,
our results are the first estimates of the double-derivative "$d$" parameters.

Although there are six Lagrangian parameters, their freedom to vary is
restricted.  The parameter $b_0$ is confined to stay between $-0.52$ and
$-0.28$ GeV $^{-1}$ as discussed above.  Requiring $a^+_{\pi N}$ and $a^0_{KN}$
to be near
zero, reduces the effective degrees of freedom to three.  Thus, a
fit to the diversified data of the $\overline{K} N$ system with only one common
range parameter (for the local potential) is remarkable.   As can be seen Table
II, the double-derivative "$d$" parameters are mostly negative.  The negative
sign is needed to cancel the quark mass "$b$" terms in $a^+_{\pi N}$ and
$a^0_{KN}$ in order to obtain the near zero experimental values of these
quantities (see Appendix II).   This same cancelation is also desirable in the
$\overline{K} N$
system, since the iterated current algebra term by itself does a fairly good
job at fitting the data.

For the $K^+ N$ system, the $I=0$ scattering length was incorporated in the fit
and is near zero.  The scattering length remains essentially zero when the
potential is iterated.  The "best fit" parameters also apply to the $K^+ N$
interaction in the $I=1$ channel, where the experimental value of the
scattering length is $-0.31 \pm 0.02$ fm \cite{dover}.  For the local
potential, the $I=1$
potential is given by Eq.(15) with the $C_{ij}$ replaced by the $C_{K^+p
\rightarrow K^+ p}$ from Appendix II.  The Born scattering length for the
"best fit" parameters gives $-0.49$ fm, and iteration of the potential
reduces this value.  Thus by choosing an appropriate range for this channel,
the
scattering length can be made equal to the experimental result.  A range
parameter of
$\alpha_{K^+ p} = 500$ MeV reproduced the experimental value of $-0.31$ fm.  We
note that this range also gives reasonable agreement with the energy dependence
of the $I=1$ s-wave phase shift as shown in Fig. 2.  Here we plot the
s-wave phase shift as a function of kaon laboratory momentum for this fit by
the solid line.  The dots correspond to the phase shift taken from Ref.
\cite{martin}.

We have investigated if there were any other satisfactory fits to the data by
using different starting points for the search.  We also tried three different
ranges with the local potential, and found one fit with $\alpha = 569$ MeV for
the $\overline{K} N$ channel, but with very similar values for the Lagrangian
parameters as in Table II.  In Fig. 2 we plot the s-wave $K^+ p$ phase shift
for this fit by the dashed line.  In this case, $\alpha_{K^+ p} = 760$ MeV, and
the energy dependence agrees even better with Martin's phases \cite{martin}.
The overall best fit used three ranges for the local potential and produced a
$\chi^2/N = 1.7$ for all the data, but again with similar values for the
Lagrangian parameters.

We note that the parameters were fitted to the data using a potential model,
and need to be interpreted in this context.  We briefly comment on possible
limitations of the potential model approach when comparing with a chiral
perturbation expansion.

When iterating the potential in the Lippmann-Schwinger equation,
four-dimensional loop integrals are not calculated in all completeness.  The
energy part of the integral is integrated out, producing a propagator
appropriate for the three remaining dimensions.   All
multiple scatterings include only "ladder" graphs.    Crossed
graphs are not included in our calculation in the iteration process, and the
meson-meson interaction is neglected.  This is expected to be a reasonable
approximation, since the energies are low, the mass of the kaon is
relatively large, and the resonance at $1405$ MeV dominates the
$\overline{K}N$ interaction at low energies.  The S-matrix we obtain in the
potential model is unitary, and has a pole at the $\Lambda (1405)$
resonance.  The Born amplitudes respect crossing symmetry, but the full
amplitudes cannot be extended too far in energy
away from the region investigated here, since neglecting crossed graphs in the
iteration of the potential violates crossing symmetry.  It is encouraging,
however, that the s-wave $K^+p$ scattering length and the energy dependence of
the phase shift are reproduced reasonably well even though only the Born
amplitudes respect crossing symmetry.

In a systematic chiral perturbation expansion amplitudes are expanded in powers
of meson energies.  The terms up to order
$q^2$ for s-wave scattering are given entirely in terms of the current algebra
term, the "$b$" and "$d$" parameters and relativistic corrections to order
$q^2$.  The $q^3$ contribution is given by single loops involving just
the current algebra term and higher order counter terms.  However, using a
potential model with a cut-off introduces an additional length scale.
This is most easily seen by examining the separable
potential for which an analytic solution exists.
For a single channel separable potential, the scattering amplitude is given by:

\begin{equation}
  f_{K^+p} = a_{Born}\biggl({\alpha^2 \over {\alpha^2 + k^2}}\biggr)^2
              \biggl[1 + a_{Born}
           {{\alpha^3 (k^2-\alpha^2)} \over
                {2 (k^2+\alpha^2)^2}} - i a_{Born} k \biggl({{\alpha^2}
                          \over
                 {\alpha^2 + k^2}}\biggr)^2 \biggr]^{-1},
\end{equation}

\noindent and the scattering length is thus:

\begin{equation}
  a = a_{Born} \biggl[1 - {\alpha \over 2} a_{Born} \biggr]^{-1} .
\end{equation}

\noindent Here $a_{Born}$ includes both the current algebra term (order $q$)
and
the tree level terms of order $q^2$.  By expanding the denominator, the next
power of "$q$" after $q^2$ is $q^2 \alpha$.  Thus, it is not possible to
compare terms from the potential model used here with those of a chiral
perturbation
expansion beyond order $q^2$.   However, one can use the above equation to
estimate how a chiral perturbation series might converge for the $K^+ p$ s-wave
scattering amplitude.  To order $q$ the scattering length is $-0.59$ fm.
Using the "best fit" parameters of the separable potential, the $q^2$ terms
contribute $-0.11$ fm to the scattering length.  Thus $a_{Born}=-0.70$ fm.
Order $q^3$ and higher terms need to be large to bring the scattering length
close to the experimental value.  For the separable potential one can calculate
the contribution of the once iterated current algebra piece, which in a chiral
perterbation expansion is the $q^3$ term.  To obtain $a_{K^+p} \simeq -0.3$ fm,
the experimental value, one needs $\alpha_{K^+p} = 667$ MeV, making the $q^2
\alpha$ term
equal to $+0.58$ fm.  Thus, we see here that the once iterated current algebra
term can be as large the order $q$ term.

Eq.(23) is instructive in comparing the difference between the $S=+1$ and
$S=-1$ $KN$ interactions.  The current algebra terms have opposite sign,
negative for $K^+p$ and positive for $K^-p$.  For the $K^+ p$ case
($a_{Born}<0$) the multiple scattering reduces the scattering length from the
Born result in accord with experiment.  Since $a_{Born}$ is roughly twice the
experimental value, a short range (large $\alpha$) is needed for the multiple
scattering to produce a large enough correction.  This short range is also
desirable in obtaining the correct energy dependence of the s-wave phase shift,
for which there is a small effective range.  For the $K^- p$ case
($a_{Born}<0$), the multiple scattering causes a zero in the denominator
producing a quasi-bound state.

The $\Lambda (1405)$ produced in the potential model derived here is a bound
state in the $\overline{K} N$ ($I=0$) channel and a resonance in the $\pi
\Sigma$ channel.  It is instructive to verify this in terms of the
$\overline{K} N$ and $\pi \Sigma$ wave functions.  In Fig.3 we plot these wave
functions, taken at the energy corresponding to the peak of the $\pi \Sigma$
mass spectrum, for the best-fit one parameter local potential (i.e. with
$\alpha = 412$ MeV).
The horizontal axis is the $\overline{K} N$ or $\pi \Sigma$ relative coordinate
$r$, and the vertical axis is the radial wave function $u(r) = r \Psi (r)$.
The root mean square radius of the $\overline{K} N$ bound state is $1.3$ fm.

Historically there has been a disagreement between the scattering length
extracted from the scattering data versus the $K^-p$ 1s atomic level shift.  In
our analysis as well, the real part has the opposite sign as the atomic level
shift results.  In Figure 4 we plot the real and imaginary part of the
scattering amplitude near and below the $K^-p$ threshold for both the local and
separable potential.  The structure below threshold is typical of a resonance.
The $K^-p$ scattering length we obtain is $(-0.97 + 1.1 \, i) $ fm, and in the
isospin limit setting $m_K = m_{\overline{K^0}}$ and $M_p = M_n$ we get
$(-0.51 +  0.89 \, i)$ fm. The experimental data is
still controversial \cite{dover}, so we did not include it in our fits.  It is
important to obtain a reliable value for the level shift, since if further
experimental analysis confirms earlier results, chiral dynamics will be in
disagreement with the data.  In general, more scattering data would also be
useful in further testing the predictions based on chiral dynamics, since the
experimental data for the scattering channels are quite old, and for some
channels not very extensive.

\section{Conclusions}

We have applied the effective chiral Lagrangian to the strangeness $S=-1$
sector of the $\overline{K}N$ interaction by constructing a pseudo-potential
from the Lagrangian, such that this potential used in the Born approximation
has
the same s-wave scattering amplitude as the chiral Lagrangian up to order
$q^2$.  This potential model approach successfully produces
the $\Lambda (1405)$ resonance just below the $K^-p$ threshold.  A number of
Lagrangian parameters were adjusted to fit the available low energy
$\overline{K}N$ data, with constraints from the $S=+1$ $KN$ system
and the isospin even $\pi N$ scattering length.  Satisfactory results were
found using both local and separable potential forms, and a good fit was
obtained
using only one common range parameter for all channels in the local potential.
The Lagrangian parameters turned out roughly the same for both potential forms
used.
To our knowledge the values obtained here are the first estimates of some of
these parameters.

\section{Acknowledgements}

During the final preparation of this paper, one of us (W.W.) enjoyed
stimulating discussions with G.E. Brown, J. Gasser, B. Holstein, D. Kaplan, Ch.
Pethick, and M. Rho at the INT, Seattle in the course of the program "Chiral
Dynamics of Hadrons and Nuclei".   One of us (P.S.) would like to thank the
Physics Department at Technische Universit\"{a}t M\"{u}nchen for the
hospitality extended to him during his sabbatical stay.

\newpage

\begin{center}
\begin{tabular}{|cc|cc|}\hline
\multicolumn{2}{|c|}{Local Potential} & \multicolumn{2}{c|}{Separable
Potential} \\
\hline
$\alpha$ & Energy & $\alpha$ & Energy \\
\hline
550 & $<$1300 & 700 & $<$1300 \\
500 &  1336 & 600 &  1367 \\
450 &  1378 & 550 &  1405 \\
400 &  1417 & 500 &  1424 \\
350 &  1431 & 450 &  1431 \\
300 &  1434 & 400 &  1434 \\
\hline
\end{tabular}
\end{center}
\vspace{.2in}
\noindent {\bf Table I}.  The energy of the $\overline{K} N$ ($I=0$)
quasi-bound state produced from the current algebra (Weinberg-Tomozawa) term
alone as a function of the range parameter $\alpha$ for both the local and the
separable potential.  The range $\alpha$ is the same for all channels.

\bigskip

\begin{center}
\begin{tabular}{|c|ccc|ccc|}
\hline
Potential & $\alpha$ (MeV) & $f$ (MeV) & $\chi^2/N$ &
 $\gamma$ & $R_c$ & $R_n$ \\
      &    &    &  (for scattering data) &   &   &   \\
\hline
Local & 395  & 90 & 1.8 & 1.34 & 0.645 & 0.15 \\
Separable & 700  & 110 & 1.8 & 1.21 & 0.652 & 0.14 \\
\hline
Exp   &   &  &  &  $2.36$ & $0.66 $ & $0.19$ \\
      &   &  &  & $\pm 0.04$ & $\pm 0.011$ & $\pm 0.015$ \\
\hline
\end{tabular}
\end{center}
\vspace{.2in}
\noindent {\bf Table II}.  Best fit results using only the current algebra term
for the local and separable potential.  The $\chi^2$ per data point is only for
the elastic and inelastic $K^-p$ scattering data.  In both cases a resonance is
formed at 1405 MeV.  The experimental value for $\gamma$ is $2.36 \pm .04$.

\bigskip

\begin{center}
\begin{tabular}{|c|cccccc|ccc|}\hline
Potential & $b_0$ & $d_0$ & $d_D$ & $d_F$ & $d_1$ & $d_2$ & $\alpha_{\Sigma
\pi}$ & $\alpha_{\Lambda \pi}$ & $\alpha_{KN}$ \\
\hline
Local   & --0.493 & --0.66 & --0.02 & --0.29 & +0.23 & --0.39 & 0.41 & 0.41 &
0.41 \\
Separable & --0.279 & --0.40 & --0.24 & --0.43 & +0.28 & --0.62 & 0.45 & 0.30 &
0.76 \\
\hline
\end{tabular}
\end{center}
\vspace{.2in}
\noindent {\bf Table III}.  Best fit values for $b_0$ and the double derivative
"$d$" coefficients for both the local and separable potentials.  For the
local potential, a common range parameter $\alpha = 0.412$ GeV for all diagonal
and off-diagonal channels produced a good fit to the data.  All Lagrangian
parameters are in GeV$^{-1}$.  The range parameters are in GeV.

\newpage

\begin{center}
\begin{tabular}{|c|cc|ccc|}\hline

Potential & $f$(MeV) & $\chi^2/N$ & $\gamma$ & $R_c$ & $R_n$  \\
\hline
Local   & 93 & 1.6 & 2.30 & 0.66 & 0.17 \\
Separable & 93 & 2.8 &  2.26 & 0.66 & 0.16 \\
\hline
Exp.      &  &  &  $2.36 $ & $0.66$ & $0.19$ \\
          &  &  & $\pm 0.04$ & $\pm 0.011$ & $\pm 0.015$ \\
\hline
\end{tabular}
\end{center}
\vspace{.2in}
\noindent {\bf Table IV}.  Best fit results of the threshold branching ratios
Eqs.(21) for both the local and the separable potentials.  The $\chi^2$ per
data point is only for the $K^- p$ elastic and inelastic scattering data.  The
parameters of the "best fit" are from Table III.

\newpage

\centerline{\bf Appendix I}

\bigskip

\renewcommand{\theequation}{A1.\arabic{equation}}
\setcounter{equation}{0}

The pseudoscalar meson field matrix $\phi$ from Eq.(4) is given by

\begin{equation}
  \phi = \sqrt{2} \left( \begin{array}{ccc} {{\eta_8} \over \sqrt{6}} +
{{\pi^0} \over \sqrt{2}} & \pi^+ & K^+ \\
 \pi^- & {{\eta_8} \over \sqrt{6}} - {{\pi^0} \over \sqrt{2}} & K^0 \\
 K^-  &  \overline{K^0} & - {2 \over \sqrt{6}} \eta_8 \end{array} \right)
\end{equation}

\noindent and the baryon matrix $B$ is

\begin{equation}
   B = \left( \begin{array}{ccc} {\Lambda \over \sqrt{6}} + {\Sigma^0 \over
   \sqrt{2}} &   \Sigma^+ &  p  \\
  \Sigma^- &  {\Lambda \over \sqrt{6}} - {\Sigma^0 \over \sqrt{2}} & n \\
  \Xi^- & \Xi^0  & - {2 \over \sqrt{6}} \Lambda  \end{array} \right)
\end{equation}

Using the Gell-Mann, Oakes, Renner (GOR) relation in the isospin limit for the
pseudoscalar mesons, the mass matrix $\chi$ is given by:

\begin{equation}
 \chi = \left( \begin{array}{ccc} m_\pi^2 & 0 & 0 \\
                                0 & m_\pi^2 & 0 \\
                                0 & 0 & 2 m_K^2-m_\pi^2  \end{array} \right)
\end{equation}

\vspace{1in}

\centerline{\bf Appendix II}

\bigskip

\renewcommand{\theequation}{A2.\arabic{equation}}
\setcounter{equation}{0}

Here we list the values of the relative coupling strengths $C_{ij}$ for the
potential of Eqs.(12,15,16) in terms of the parameters of the chiral
Lagrangian.
Channel one corresponds to the $\pi^+ \Sigma^-$ state, channel two to $\pi^0
\Sigma^0$, channel three to $\pi^- \Sigma^+$, channel four to $\pi^0 \Lambda$,
channel five to $K^- p$, and channel six to the $\overline{K^0} n$ state.

\begin{eqnarray}
   C_{11} & = & -E_\pi' + (F^2 + {D^2 \over 3})S_{\pi \pi} + 4 m^2_{\pi}
                          (b_D + b_0) -
               E^2_\pi  (2 d_D + 2 d_0 + d_1) \nonumber \\
   C_{12} & = & -E_\pi' + {D^2 \over 3} S_{\pi \pi} - F^2 U_{\pi \pi} -
               E^2_\pi (d_1+d_2) \nonumber \\
   C_{13} & = & -2 E^2_\pi (d_1 + d_2) + ({D^2 \over 3} - F^2)
                    (S_{\pi \pi} + U_{\pi \pi}) \nonumber \\
   C_{14} & = & {DF \over \sqrt{3}}(S_{\pi \pi} - U_{\pi \pi}) \nonumber \\
   C_{15} & = &  -(F^2+{D^2 \over 3}) {S_{\pi K} \over 2} +
                (D^2-F^2) {U_{\pi K} \over 2} -E_K E_\pi (d_1 + d_2) \nonumber
\\
   C_{16} & = & -{1 \over 4}(E_\pi' + E_K')+({F^2 \over 2} - DF - {D^2 \over
6})
                        S_{\pi K}  \nonumber \\
          &   & - (m^2_\pi + m^2_K)(b_F - b_D) - E_\pi E_K (d_D - d_F + d_1)
\nonumber \\
   C_{22} & = & C_{11} - C_{12} + C_{13} \nonumber \\
   C_{23} & = & C_{12} \nonumber \\
   C_{24} & = & 0 \nonumber \\
   C_{25} & = & {1 \over 2} (C_{15} + C_{16}) \nonumber \\
   C_{26} & = & C_{25} \nonumber \\
   C_{33} & = & C_{11} \\
   C_{34} & = & -C_{14} \nonumber \\
   C_{35} & = & C_{16} \nonumber \\
   C_{36} & = & C_{15} \nonumber \\
   C_{44} & = & {D^2 \over 3}(S_{\pi \pi} + U_{\pi \pi})+ 4 m_\pi^2
                   (b_0+{b_D \over 3})
             - E_\pi^2(2 d_0 + {{2 d_D} \over 3} + {d_2 \over 3})\nonumber \\
   C_{45} & = & -{\sqrt{3} \over 8} (E_\pi' + E_K')+
                            {D \over 2 \sqrt{3}}(D-F)S_{\pi K} -
            {\sqrt{3} \over 4}(D+F)({D \over 3} + F) U_{\pi K} \nonumber \\
          &   & - {1 \over 2 \sqrt{3}}(m^2_\pi + m^2_K)
               (b_D + 3b_F) + {1 \over 2 \sqrt{3}} E_\pi E_K
                (d_D+3d_F-d_2) \nonumber \\
   C_{46} & = & -C_{45} \nonumber \\
   C_{55} & = & -E_K' +(F^2+{D^2 \over 3})S_{KK} + 4 m^2_K (b_D+b_0) -
              E^2_K (2 d_D + 2d_0+d_1) \nonumber \\
   C_{56} & = & -{E_K' \over 2} +({F^2 \over 2}+DF-{D^2 \over 6})S_{KK} + 2
m^2_K
                   (b_D+b_F) - E^2_K (d_D+d_F+d_1) \nonumber \\
   C_{66} & = & C_{55} \nonumber
\end{eqnarray}

\noindent where $E_\pi$ and $E_K$ denote the pion and kaon energy respectively
in the center-of-mass frame.  The pion and kaon masses are labeled as $m_\pi$
and $m_K$.  The quantities labeled $E_\pi'$ and $E_K'$ include the
relativistic correction to the current algebra piece to order $q^2$ and are:

\begin{equation}
  E_\alpha' = E_\alpha + {{E_\alpha^2 - m_\alpha^2} \over {2 M_0}}
\end{equation}

\noindent The quantities $S_{\alpha \beta}$ and $U_{\alpha \beta}$ are defined
as:

\begin{eqnarray}
  S_{\alpha \beta} &=& {{E_\alpha E_\beta} \over {2 M_0}} \nonumber \\
  U_{\alpha \beta} &=& {1 \over {3 M_0}} \biggl( 2m^2_\alpha + 2m^2_\beta +
                   {{m^2_\alpha m^2_\beta}
           \over {E_\alpha E_\beta}} - {7 \over 2} E_\alpha E_\beta \biggr)
\end{eqnarray}

\noindent where the index $\alpha$ (or $\beta$) stands for pion or kaon.  These
terms arise
from the Born graphs involving the axial vector couplings which have an octet
baryon exchanged in the "s" or "u" channel.  $M_0$ is the
baryon octet mass in the chiral limit which we take as $910 \, MeV$.  Note that
$C_{ij} = C_{ji}$.

The ${1 \over E}$ terms in $U_{\alpha \beta}$ deserve some explanation.  Baryon
exchange in the u-channel leads to logarithmic singularities in the partial
wave amplitudes below threshold.  In the $\pi N$ case it gives rise to the
so-called short nucleon cut \cite{hoehler} extending from $s=(M_N - {{m^2_\pi}
\over M_N})^2$ to
$s=M_N^2 + 2 m^2_\pi$.  In the heavy mass limit, both logarithmic branch points
in the $E_\pi$-plane coalesce and give rise to a pole singularity.  The
presence
of subthreshold singularities restricts the kinematic range for the potential
approach to be applied.  In this analysis the energies are well above these
singularities, the nearest coming from $\pi \Sigma \rightarrow \overline{K}
 N$ is
located at $\sqrt{s} = 1.24$ GeV ($E_\pi = 58$ MeV).  Even if one used the full
expression for the u-channel exchange (producing logarithmic singularities)
instead of the heavy mass limit to order $q^2$ (producing pole singularities
very close to the logarithmic ones) the results of the analysis would not
change.  In fact,  when we neglected the s- and u- channel baryon exchange
amplitudes entirely, we obtained a good fit to the data with similar values for
$b_0$ and the "$d$" parameters as in Table II.

The potential strength for the $I=1$ channel for $K^+N$ scattering (i.e. $K^+p$
scattering is given by:

\begin{eqnarray}
   C_{K^+p \rightarrow K^+p} &=& E_K + {{E^2_K-m^2_K} \over {2M_0}} + (F^2+{D^2
   \over 3})U_{KK} \nonumber \\
            & & +4m^2_K(b_0+b_D) - E^2_K(2d_D+2d_0+d_1)
\end{eqnarray}

In terms of the Lagrangian parameters, the $\pi N$ isospin-even scattering
length to order $q^3$ is given by \cite{loop}:

\begin{eqnarray}
  a^+_{\pi N} &=& {{1} \over {4 \pi (1+m_\pi /M_N)}}\biggl[ {{m^2_\pi} \over
            {f^2}} \biggl(-2b_D - 2b_F - 4b_0 + d_D + d_F + 2 d_0 -
                   {{g^2_A} \over {4 M_N}} \biggr) \nonumber \\
             & & +  {{3 g^2_A m^3_\pi} \over {64 \pi f^4}}  \biggr]
\end{eqnarray}

\noindent where $g_A$ is the axial vector coupling constant.  The experimental
value of the even $\pi N$ scattering length is $-0.012 \pm 0.006$ fm
\cite{koch}.  The $I=0$ for
the strangeness $S=+1 \; KN$ interaction is given to order $q^2$ (i.e. tree
level) as:

\begin{equation}
  a^0_{KN} = {{1} \over {4 \pi (1 + m_K /M_N)}} {{m^2_K} \over {f^2}}
                \biggl(4 b_0 - 4b_F + 2 d_F - 2 d_0 + d_1 +
                 {D \over M_N}(F-{D \over 3}) \biggr)
\end{equation}

\noindent  The experimental value for this quantity is $-(0.02 \pm 0.04)$ fm
\cite{dover}.

\newpage

\centerline{Figure Captions}

\bigskip

Figure 1.  Calculated cross sections and resonance spectrum compared with
experimental data \cite{ciborowski,evans,humphrey,kim,sakitt}: a) $K^-p$
elastic scattering, b) $K^-p \rightarrow K^0n$,
 c) $K^-p \rightarrow \pi^0 \Lambda$, d) $K^-p \rightarrow \pi^+ \Sigma^-$,
e) $K^-p \rightarrow \pi^0 \Sigma^0$, f) $K^-p \rightarrow \pi^- \Sigma^+$,
and g) the $\Sigma \pi$ mass spectrum.  The solid curves correspond to the best
fit parameters for the local potential, and the dashed curve to the separable
potential.

\bigskip

Figure 2.  The $I=1$ $K^+ N$ (i.e. $K^+ p$) phase shift as a function of kaon
laboratory momentum.  The solid line corresponds to the best-fit using only one
common range parameter ($\alpha = 412$ MeV) for the local potential.  A value
of  $\alpha_{K^+ p} = 500$ MeV reproduces the experimental value of the $K^+ p$
scattering length.  The dashed line corresponds to a three parameter fit for
the local potential with a larger value for $\alpha_{KN}$ as mentioned in the
text.  In this case, $\alpha_{K^+ p}$ is equal to $760$ MeV.  The dots are from
the phase shift analysis of Martin \cite{martin}.

\bigskip

Figure 3.  Radial wave functions $u(r)=r \Psi (r)$ for the $\overline{K} N$ and
$\pi \Sigma$ $I=0$ channel at $\sqrt{s} = 1405$ MeV are plotted as a function
of the relative coordinate $r$.

\bigskip

Figure 4.  The $I=0$ and $I=1$ scattering amplitudes for the $\overline{K} N$
system at and below the $K^- p$ threshold.

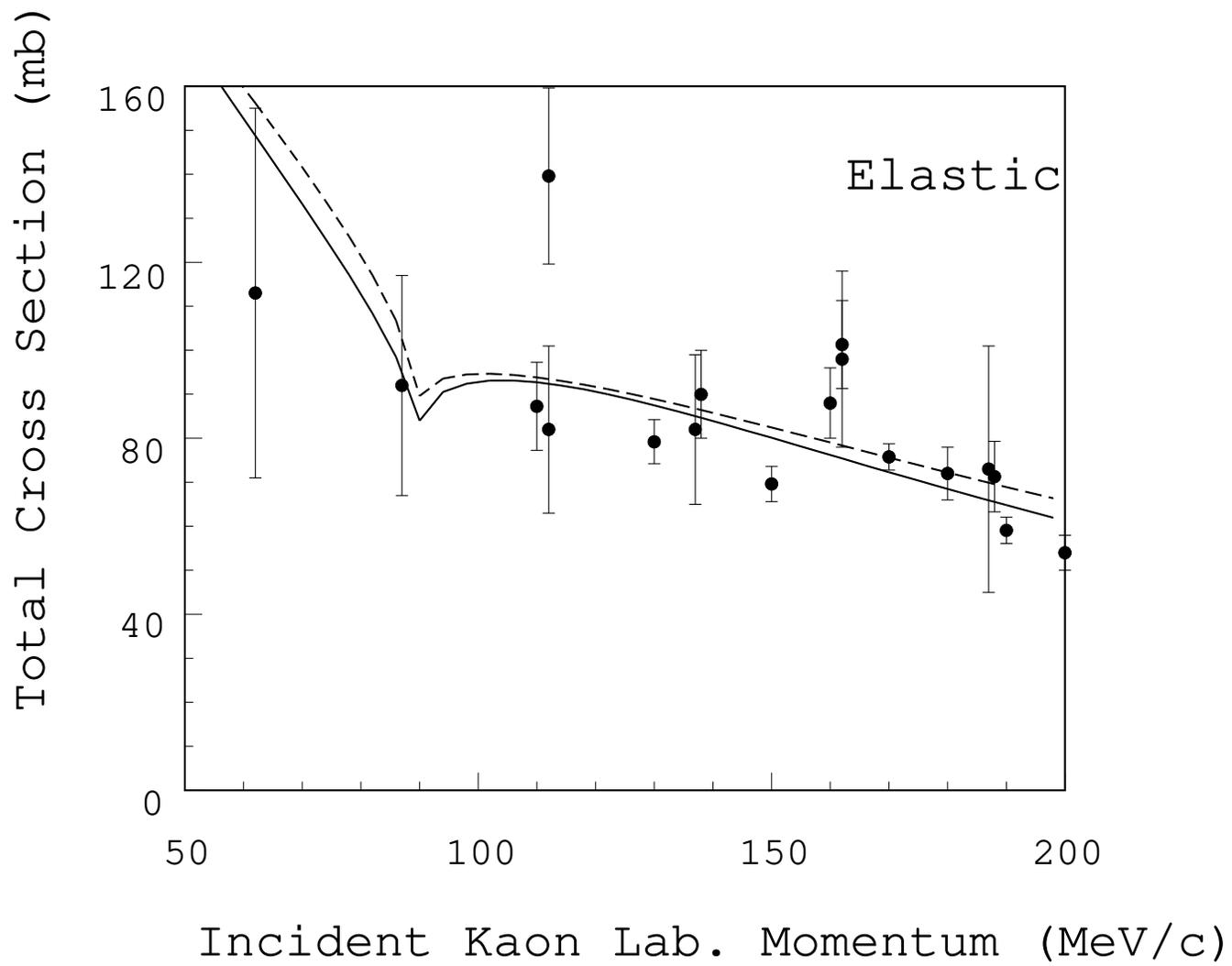
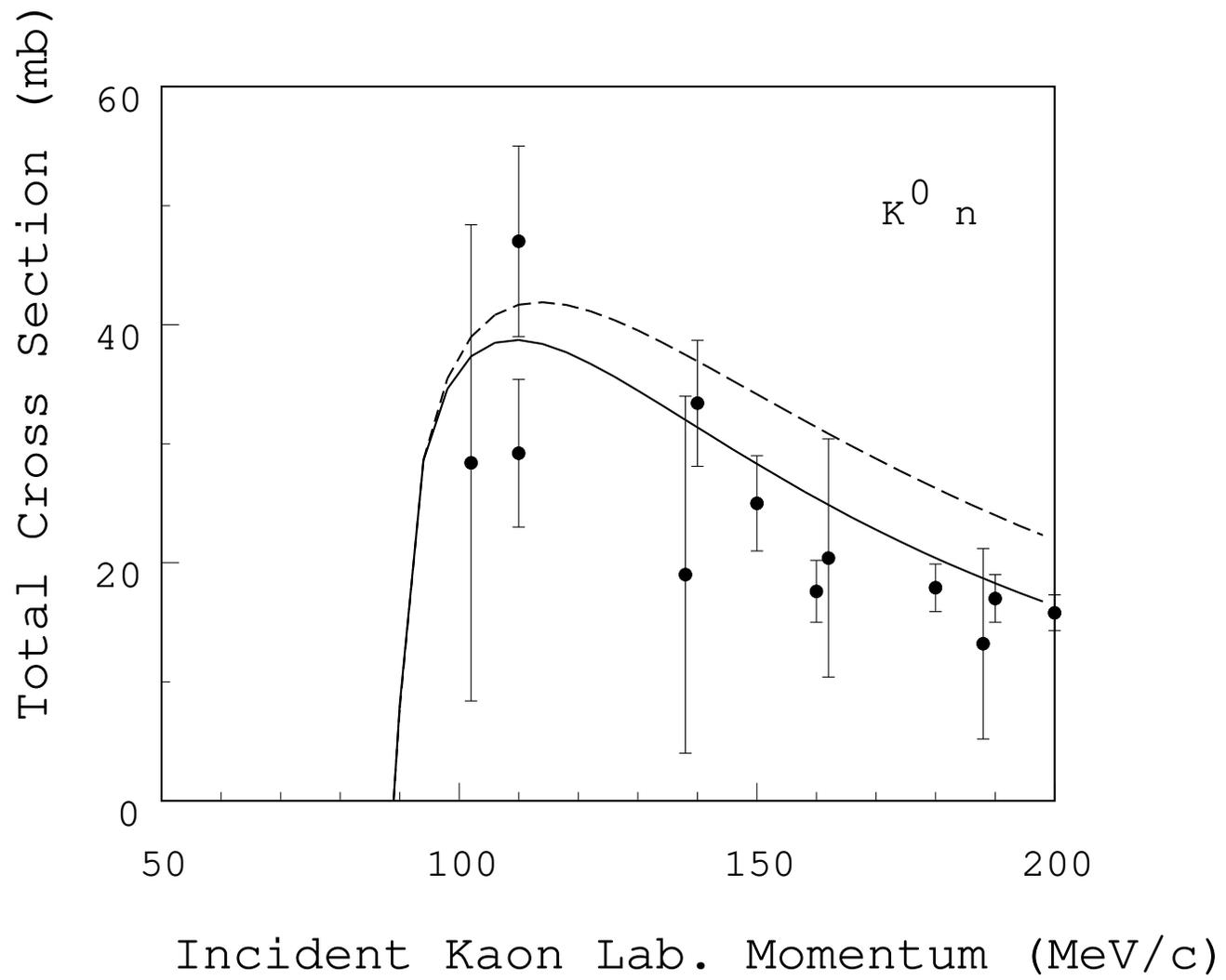
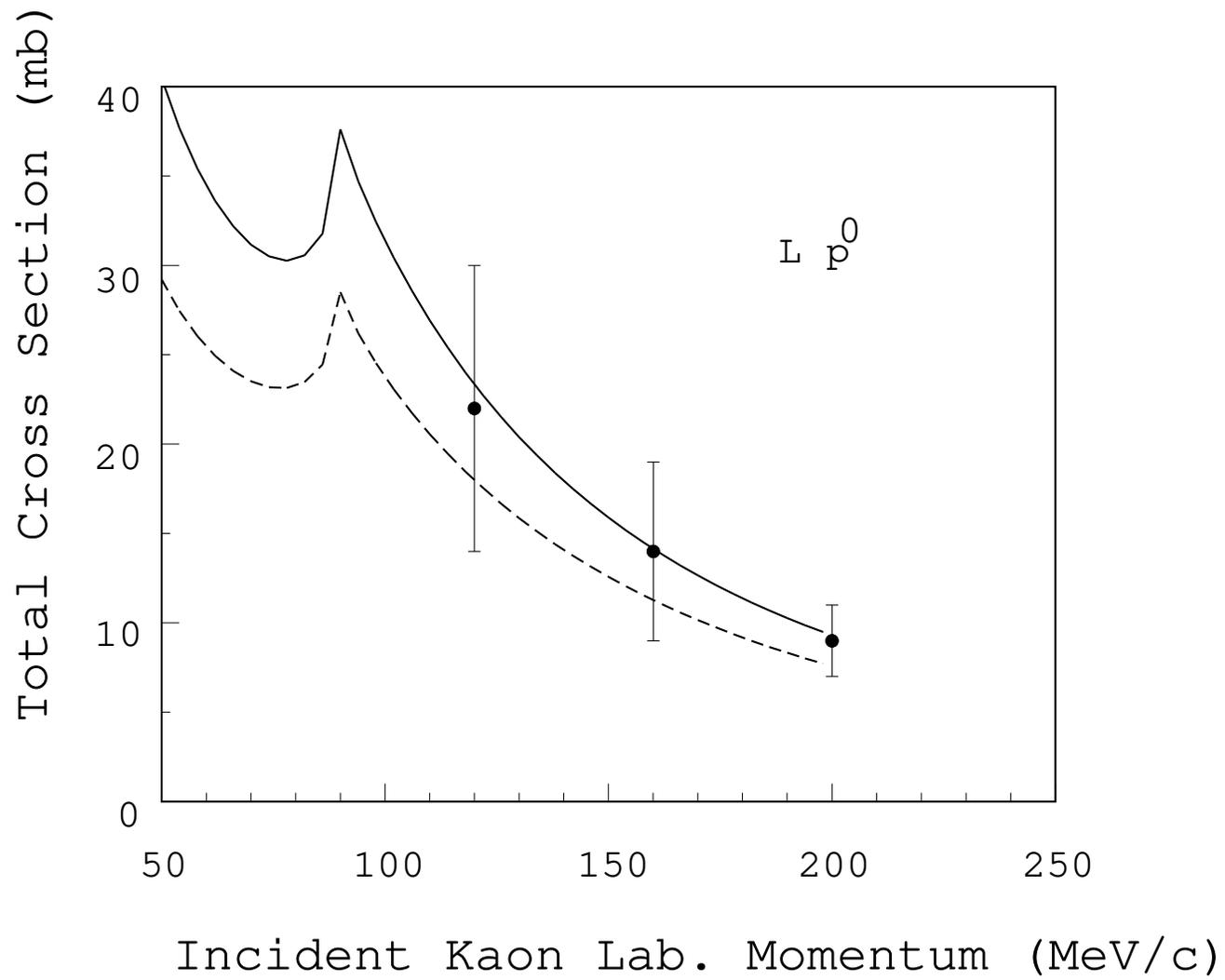
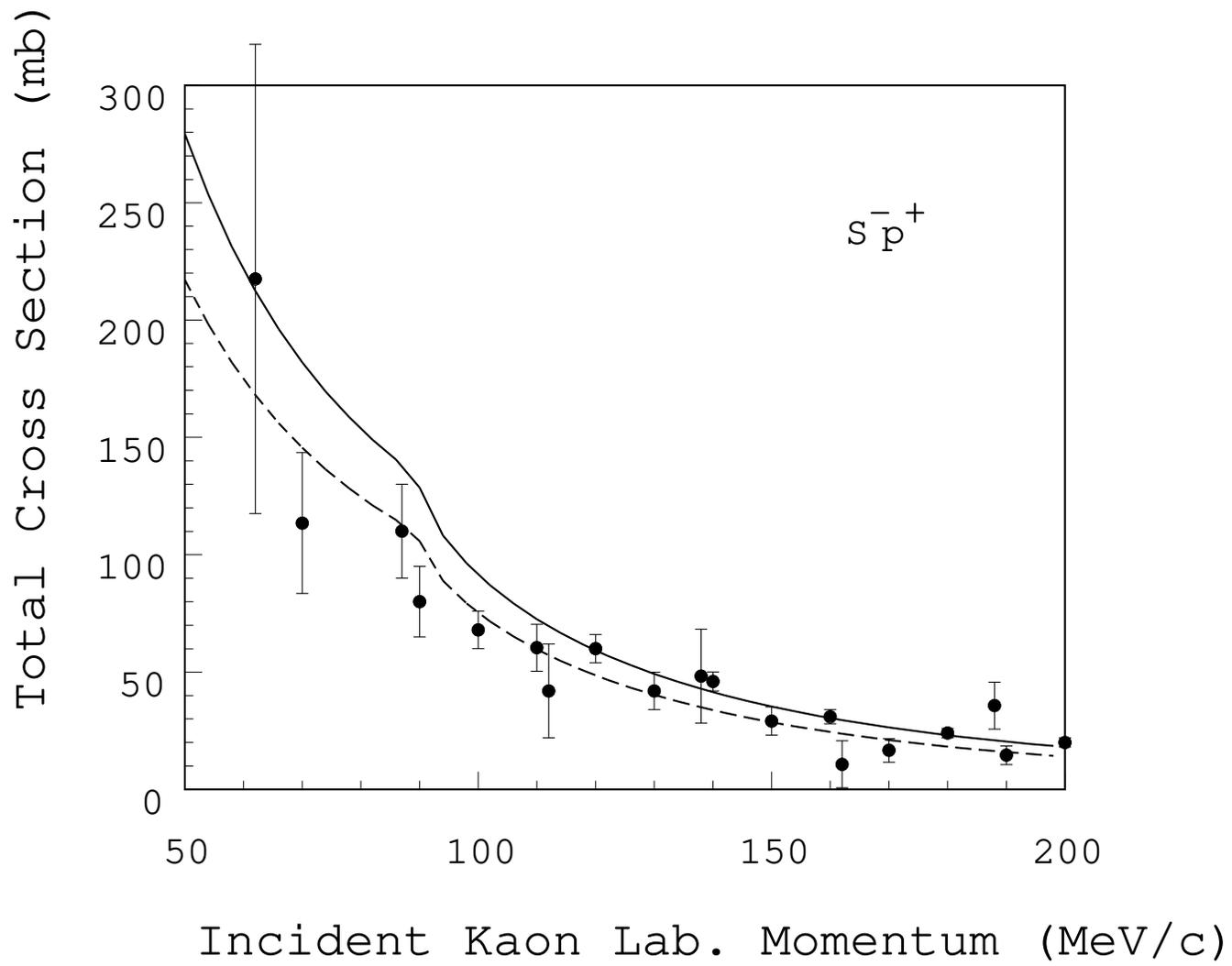
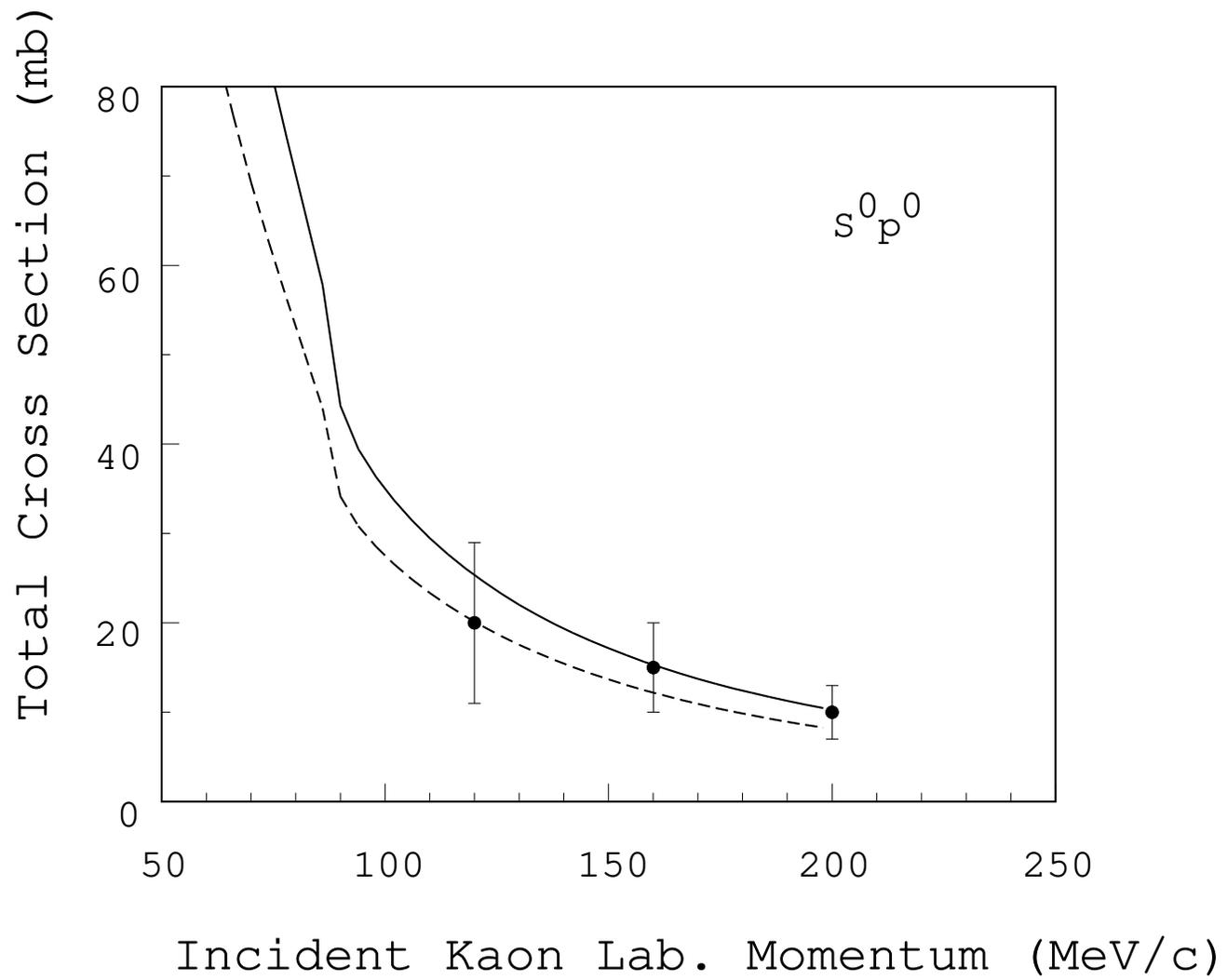
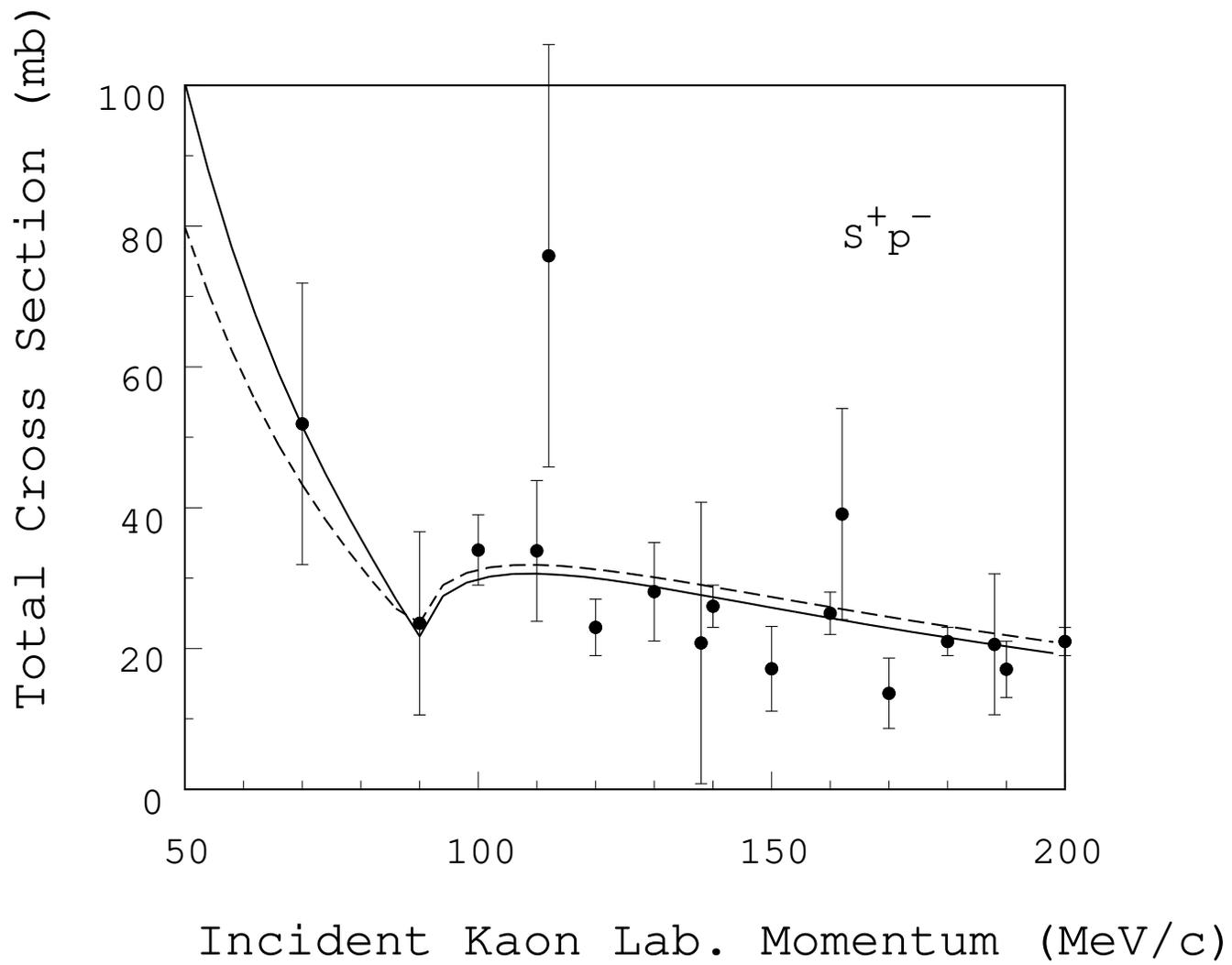
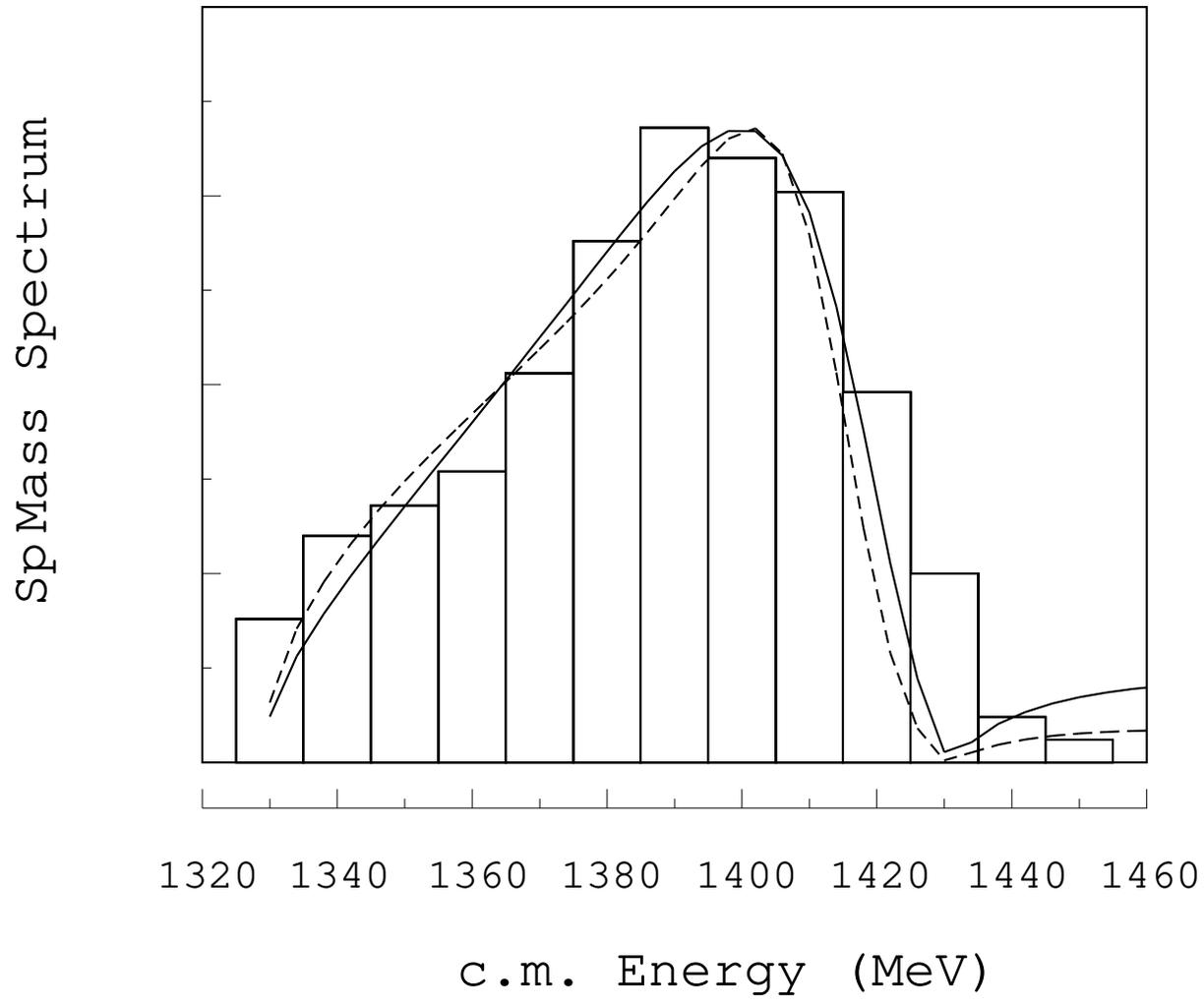
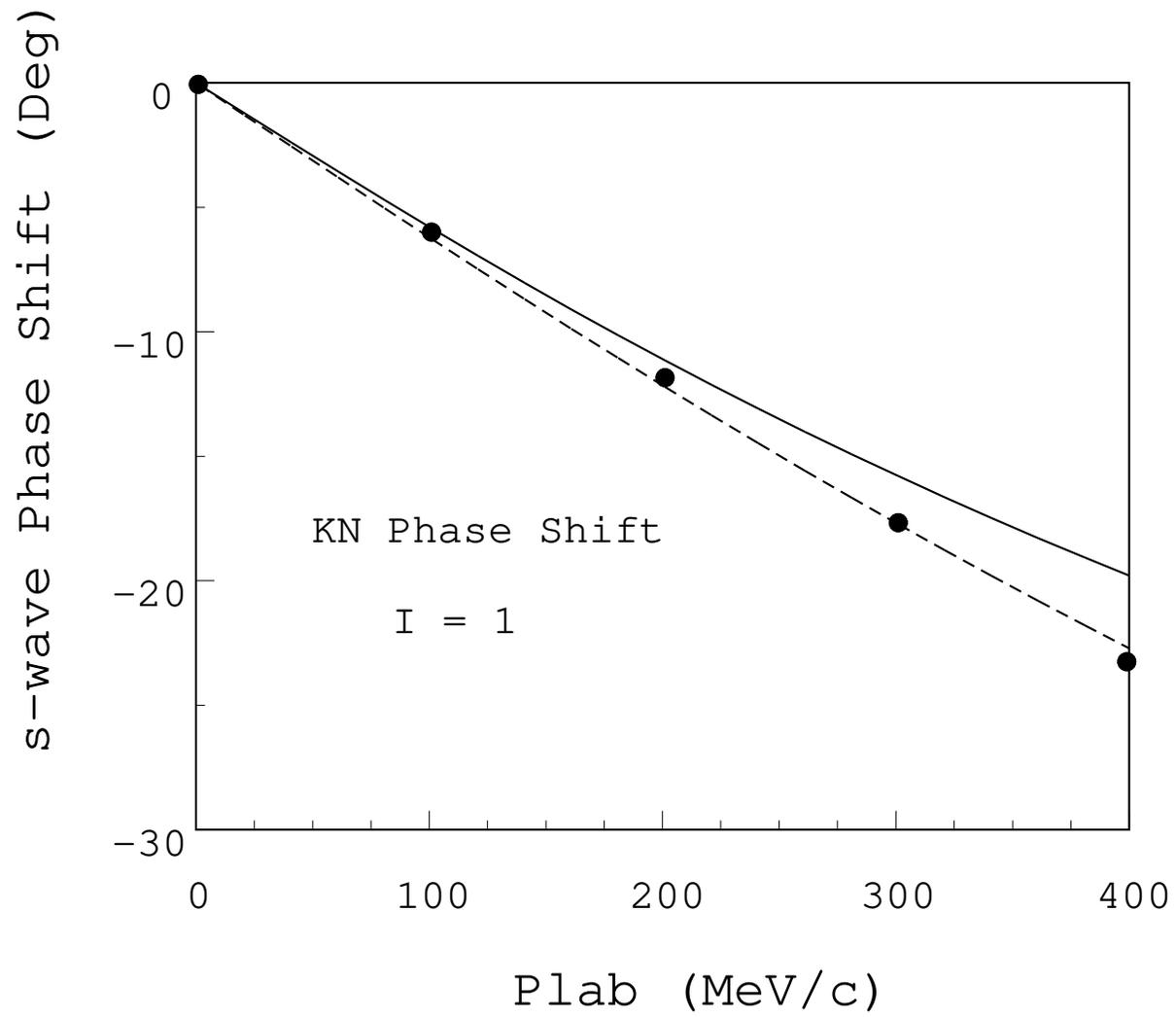
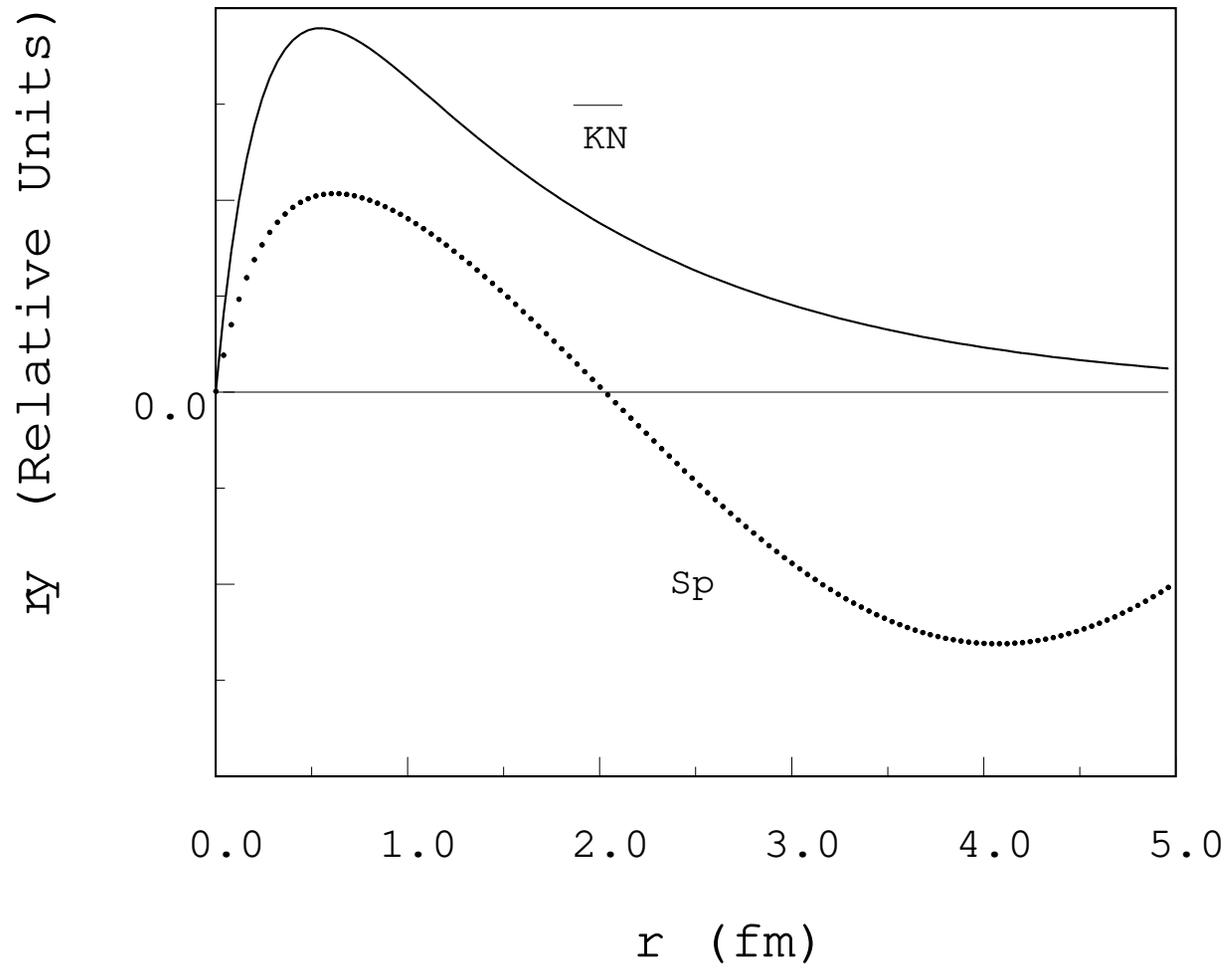
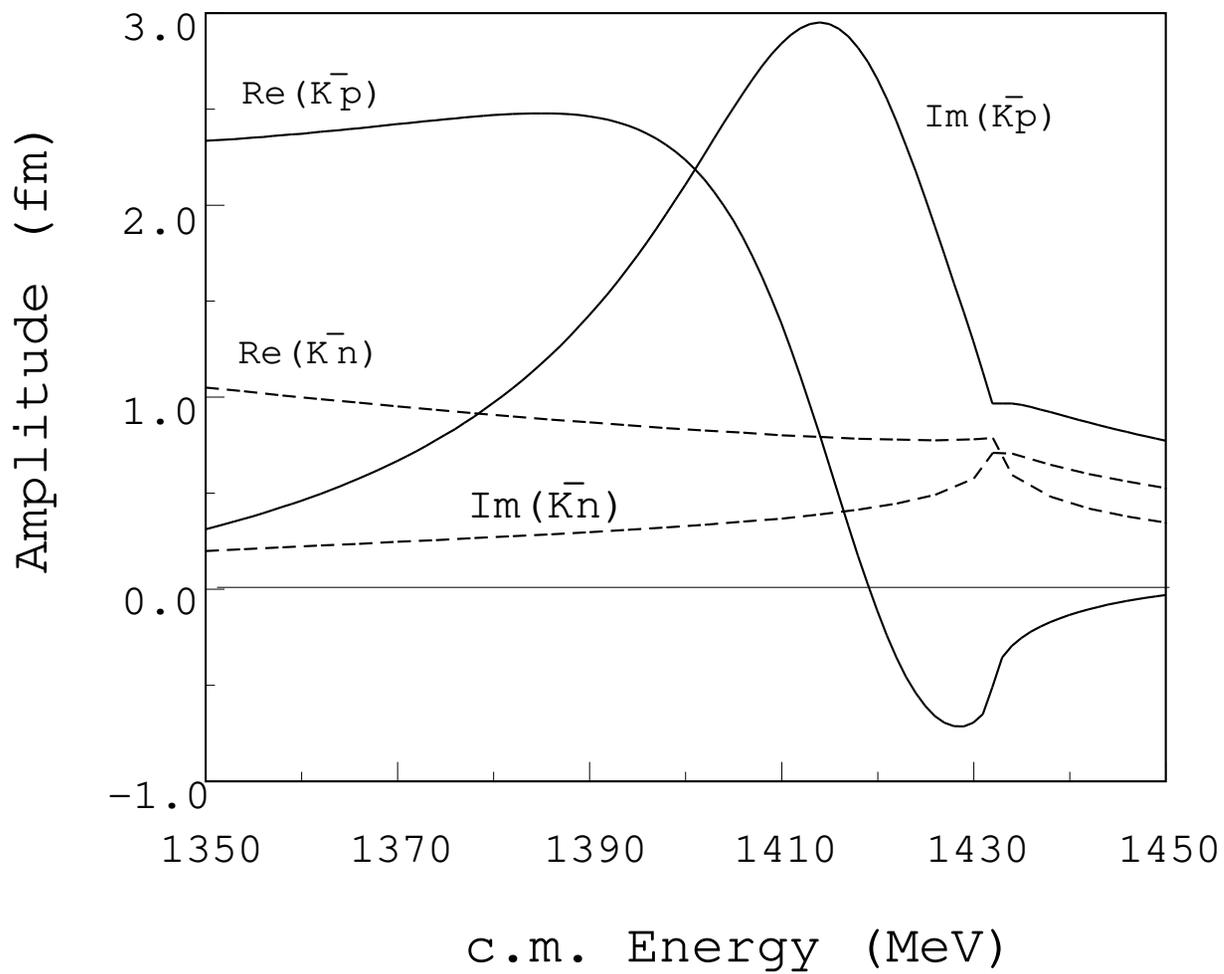
\end{document}